\documentclass[a4paper, amsfonts, amssymb, amsmath, reprint, showkeys, nofootinbib, oneside, twocolumn]{revtex4-1}
\usepackage{graphicx}  
\usepackage{dcolumn}   
\usepackage{bm}        
\usepackage{amssymb}   
\usepackage{dutchcal}
\usepackage[toc,page]{appendix}
\usepackage{amsmath}
\usepackage{braket}
\usepackage{mathtools}
\begin{document}

\title{Time-dependent conserved operators for autonomous systems and quantization of resistance}
\author{Jorge A. Lizarraga}
\email{jorge\_lizarraga@icf.unam.mx\\}
\affiliation{Instituto de Ciencias F\'isicas, Universidad Nacional Aut\'onoma de M\'exico, 62210, Cuernavaca, M\'exico}

\date{\today}

\begin{abstract}
Two systems for a charged particle are studied using an adaptation of Lewis and Riesenfeld invariant method. The first system involves a particle under the effect of a constant electric field, and the second system adds a parallel magnetic field. In both cases, time-dependent conserved operators are identified, which can be used to derive time-dependent wave functions for the Schr\"odinger equation, where the time variable is not separable from the space coordinates. These conserved operators are also used to construct unitary operators, which define the symmetries of the systems. Finally, it is shown that the invariance of the wave function under the action of these unitary operators leads to the quantization of resistance equal to an integer number times the Klitzing's constant.
\end{abstract}
\maketitle

\section{Introduction}

Conserved operators in dynamical systems have proven to be powerful tools in quantum mechanics for identifying suitable bases and understanding the structure of quantum states. A classical example is the central force problem, where the conservation of angular momentum allows one to determine the eigenfunctions associated with the azimuthal and polar angles \cite{de2014introduccion}. Another well-known case involves a charged particle in a uniform magnetic field. By choosing Landau’s gauge, the linear momentum becomes a conserved quantity, leading to the so-called Landau ansatz. The corresponding eigenfunctions are products of plane waves and harmonic oscillator wave functions \cite{landau2013quantum}.

More recently, it has been shown that for a charged particle in a constant electromagnetic field, a complete description of the system requires not only Landau’s ansatz but also a time-dependent conserved operator \cite{lizarraga2024quantization}. In that work, it was demonstrated that requiring the wave function to be invariant under a unitary transformation generated by this conserved operator leads to the quantization of resistivity for a single particle. Interestingly, the resulting expression closely resembles the quantized resistance observed experimentally by Störmer \cite{tsui1982two, stormer1999nobel, stormer1999fractional}. Moreover, the invariance condition yields two distinct quantization rules: one for the magnetic flux (also derived through Landau's ansatz \cite{landau2013quantum}) and another associated with the electric field, expressed in Eq. (\ref{quan_elect_cond}). The latter serves as the primary motivation for the present study.

In this work, we analyze two autonomous quantum systems that admit time-dependent conserved operators and naturally lead to the quantization condition for the electric field. The first system describes a particle subjected to a constant electric field, while the second considers a particle in a constant electromagnetic field with parallel electric and magnetic components. The linear potential involved in the first case is particularly relevant in studies of energy level shifts (Stark effect) \cite{robinett2009stark, matin2014solving, ito2024exact, song2024theoretical}, and time-dependent solutions have previously been obtained using the generalized Baker-Campbell-Hausdorff formula for arbitrary initial conditions \cite{soto2015solution}. However, the approach adopted here is based on the conserved quantities of the system, employing the Lewis–Riesenfeld invariant method which has proven to be a powerful tool to address time-dependent Hamiltonians \cite{FringTenney,patra2023two,ramos2018lewis,fring2022lewis,patra2024tuning,lawson2018lewis,mana2020time,koussa2020pseudo}. Although originally developed for non-autonomous systems \cite{lewis1969exact, guedes2001solution, guerrero2015lewis}, we show that this method can also be effectively applied to time-independent (autonomous) Hamiltonians.

The second configuration has been explored in various contexts. Classically, it has been used to study photodetachment from negative ions and the spatial distribution of monoenergetic electrons \cite{yukich2003observed, bracher2006electron, gao2007electron}; semiclassically, it has played a key role in photoionization microscopy experiments \cite{wang2010photoionization, deng2016photoionization}; and in fully quantum treatments, it has been employed to investigate resonances in non-hydrogenic Rydberg atoms, hydrogen ground and excited states, lithium atoms, and atomic photoionization spectra \cite{seipp1996atomic, sahoo2001resonance, johnson2001photoionization, suno1999circular}.

Beyond these specific applications, the central result of this work is the demonstration that time-dependent wave functions in the studied systems require the resistance to be quantized, provided they remain invariant under a unitary transformation generated by a time-dependent conserved operator. In particular, the resistance takes the form $R=hn/q^{2}$, where $n\in\mathbb{N}$, making it quantized in units of the von Klitzing constant $h/q^{2}$ \cite{klitzing1980new, von2017quantum}. Notably, this expression mirrors the quantized resistance measured in the experiments by Störmer that led to the discovery of the Fractional Quantum Hall Effect (FQHE) \cite{tsui1982two}. However, in contrast to the FQHE—which arises from electron–electron interactions in a two-dimensional electron gas under strong magnetic fields—the quantization presented for the first system emerges from a purely single-particle system constrained to one dimension and influenced only by a constant electric field. This suggests that similar forms of resistance quantization can arise from fundamentally different physical mechanisms. In the second system considered, although it is two-dimensional, the Hamiltonian still describes a single-particle scenario, and the same quantized expression is recovered—indicating that the quantization persists along the direction in which the electric field is applied. Interestingly, recent studies have shown that single-particle spectral features can serve as experimental signatures of Fractional Quantum Anomalous Hall Effect (which is the observation of FQHE without a magnetic field) states in van der Waals heterostructures, highlighting the intriguing possibility that FQHE signatures may be accessible through single-particle measurements \cite{pichler2025single}.

\section{invariant method}
In this section, the invariant method adapted to autonomous systems with electromagnetic field is reviewed briefly. The systems addressed in this work are described by the non-relativistic spinless particle in an constant electromagnetic field such that the Hamiltonian, written in centimetre-gram-second (CGS) units, is
\begin{equation}\label{H}
{\hat H}=\frac{1}{2m}\left({\bf \hat P}-\frac{q}{c}{\bf A}\right)^2+qU,
\end{equation}
where $m$ is the mass of the particle, $q$ is the charge of the particle, $c$ is the speed of light, ${\bf\hat P}=({\hat p}_{x},{\hat p}_{y},{\hat p}_{z})=-i\hbar\nabla$ is the momentum operator, ${\bf A}=(A_{x},A_{y},A_{z})$ is the magnetic vector potential such that the magnetic field is given by its curl, that is ${\bf B}=\nabla\times{\bf A}$ and $U=U(x,y,z)$ is the electric potential such that the electric field is given by the negative of its gradient, ${\bf E}=-\nabla U$. Given our interest in identifying time-dependent conserved operators, it is important to demonstrate that they can be used to find solutions to the Schr\"odinger equation. The total variation of an operator, ${\hat f}$, is given by the Heisenberg equation
\begin{equation}\label{dfdt}
\frac{d{\hat f}}{dt}=\frac{1}{i\hbar}[{\hat f},{\hat H}]+\frac{\partial {\hat f}}{\partial t}.
\end{equation}
When the above expression is equals to zero, one can say that the operator is conserved, that is
\begin{equation}\label{dfdt0}
\frac{1}{i\hbar} [{\hat f},{\hat H}]+\frac{\partial {\hat f}}{\partial t}=0 .
\end{equation}

Here, ${\hat f}$ is a time-dependent operator which can be used to write down an eigenvalue equation
\begin{equation}\label{eigenvalue_f}
{\hat f}\psi=\lambda\psi,
\end{equation}
where $\lambda$ is a scalar. Then, applying Eq.(\ref{dfdt0}) to $\psi$ we can write  
\begin{equation}\label{expr1}
{\hat f}({\hat H}\psi)-{\hat H}({\hat f}\psi)+i\hbar\frac{\partial {\hat f}}{\partial t}\psi=0,
\end{equation}
using Eq.(\ref{eigenvalue_f}) and $\partial_{t}({\hat f})\psi=\partial_{t}({\hat f}\psi)-{\hat f}\partial_{t}\psi$ the above expression can be rewritten as follows 
\begin{equation}\label{}
({\hat f}-\lambda)\left({\hat H}\psi-i\hbar\frac{\partial\psi}{\partial t}\right)=0 .
\end{equation}
The above equality can be satisfied for the following reasons: 1) because $\hat{H}\psi-i\hbar\partial_{t}\psi$ is an element of the kernel of the operator $\hat{f}-\lambda$ defined as the set of all square-integrable functions in $\mathbb{R}^{3}$ that satisfies the previous expression,
\begin{equation}\label{}
\text{Ker}(\hat{f}-\lambda)=\left\{\Phi\in L^{2}(\mathbb{R}^{3})|(\hat{f}-\lambda)\Phi=0\right\}.
\end{equation}
2) Because $\psi$ satisfies the Schr\"odinger equation 
\begin{equation}\label{}
{\hat H}\psi=i\hbar\frac{\partial\psi}{\partial t} .
\end{equation}

\section{One-dimensional constant electric field}
The simplest example where this approach can be used is when we work with a constant force in one dimension. In this case, the force is produced by the electric field, and the electric potential is given by $U(x)=-{\cal E}x$ where ${\cal E}$ is a constant representing the electric field intensity. The time-dependent Schrödinger equation is
\begin{equation}\label{schro_time_electric}
i\hbar\frac{\partial\psi}{\partial t}=\frac{1}{2m}{\hat p_{x}}^{2}\psi-q{\cal E}x\psi,
\end{equation}
for this case, it is known that the solutions for the stationary equation are the Airy function \cite{arfken}. However, using Eq.(\ref{dfdt}) one can realize that this system has the following conserved operators
\begin{equation}\label{f}
{\hat f}={\hat p}_{x}-q{\cal E}t,
\end{equation}
\begin{equation}\label{E}
{\hat E}=i\hbar\frac{\partial }{\partial t}.
\end{equation}
Therefore one can write down the eigenvalue equation for the operator (\ref{f}), 
\begin{equation}
    \hat{p}_{x}\psi-q{\cal E}t\psi=\lambda \psi,
\end{equation}
where $\lambda$ are the eigenvalues of the operator. These eigenvalues can be interpreted as a variation of time, the above equation can be rewritten as
\begin{equation}
    \hat{p}_{x}\psi=q{\cal E}\left(t+\frac{\lambda}{q{\cal E}}\right)\psi,
\end{equation}
and defining the variation as $\delta t=\pm\lambda/q{\cal E}$ having second as units, in this particular case the negative sign was chose and the results do not depend on the selection of this sign. Then, the eigenvalue equation reads as follows
\begin{equation}\label{eigen_value_electric}
{\hat p}_{x}\psi-q{\cal E}t\psi=-q{\cal E}\delta t\psi,
\end{equation}
Solving the above equation, it can be found that
\begin{equation}\label{}
\psi(x,t)={\cal C}(t)\exp\left(i\frac{q{\cal E}}{\hbar}(t-\delta t)x\right),
\end{equation}
where ${\cal C}(t)$ is a function that depends only on time. To determine this function, we substitute the above wavefunction into Eq.(\ref{schro_time_electric}) to obtain the following equation
\begin{equation}\label{schro_C}
i\hbar\frac{\partial{\cal C}}{\partial t}=\frac{q^{2}{\cal E}^{2}}{2m}(t-\delta t)^{2}{\cal C},
\end{equation}
having the following solution
\begin{equation}\label{}
{\cal C}(t)=\exp\left(-i\frac{q^{2}{\cal E}^{2}}{6m\hbar}(t-\delta t)^{3}\right),
\end{equation}
and the solution can be written as
\begin{equation}\label{psi_xt_electric}
\psi(x,t)=\frac{1}{L}\exp\left(-i\frac{q^{2}{\cal E}^{2}}{6m\hbar}(t-\delta t)^{3}+i\frac{q{\cal E}}{\hbar}(t-\delta t)x\right),
\end{equation}
where the normalization constant was calculated over the finite interval $x\in[-L/2,L/2]$, which can be interpreted as placing the particle within a uniform electric field generated by a parallel-plate capacitor. Although the potential formally extends over the entire real line, this restriction provides a physically motivated and mathematically convenient framework that avoids divergences in the normalization process.
At this point, one can search for the simplest form of the solution, in a sense that the above wave function is just a unitary transform of the new function. The conserved operators in Eq.(\ref{f}) and Eq.(\ref{E}) can be used to define a couple of unitary operators 
\begin{equation}\label{Uf}
{\hat U}_{x}=\exp\left(-i\frac{\delta x}{\hbar}({\hat p}_{x}-q{\cal E}t)\right),
\end{equation}
\begin{equation}\label{UE}
{\hat U}_{t}=\exp\left(i\frac{\delta t}{\hbar}{\hat E}\right),
\end{equation}
where $\delta x$ is a constant having units of length. These operators define the symmetries of the system such that the Schrödinger equation remains invariant under a unitary transformation, i.e.,
\begin{equation}\label{UHEU}
{\hat H}-{\hat E}={\hat U}_{x}^{\dagger}({\hat H}-{\hat E}){\hat U}_{x}={\hat U}_{t}^{\dagger}({\hat H}-{\hat E}){\hat U}_{t}.
\end{equation}
Note that the wave function in Eq.(\ref{psi_xt_electric}) can be written as
\begin{equation}\label{}
\psi(x,t)={\hat U}_{t}\frac{1}{L}\exp\left(-i\frac{q^{2}{\cal E}^{2}}{6m\hbar}t^{3}+i\frac{q{\cal E}}{\hbar}tx\right),
\end{equation}
which satisfies 
\begin{equation}\label{}
{\hat H}\psi(x,t)-{\hat E}\psi(x,t)=0.
\end{equation}
multiplying the above equality by ${\hat U}_{t}^{\dagger}$, using Eq.(\ref{UHEU}) and the fact that ${\hat U}_{t}$ is unitary, we have 
\begin{equation}\label{}
{\hat H}\varphi(x,t)-{\hat E}\varphi(x,t)=0,
\end{equation}
where 
\begin{equation}\label{fundamentalsol1}
\varphi(x,t)={\hat U}_{t}^{\dagger}\psi=\frac{1}{L}\exp\left(-i\frac{q^{2}{\cal E}^{2}}{6m\hbar}t^{3}+i\frac{q{\cal E}}{\hbar}tx\right).
\end{equation}

And this is the simplest form of the searched wave function. However, there is still another important characteristic of this solution to be analyzed, and is its degeneracy. Since the operator in Eq.(\ref{E}) is conserved, $[{\hat E},{\hat H}]=0$, the following equality is satisfied
\begin{equation}\label{}
{\hat E}({\hat H}\varphi)={\hat H}({\hat E}\varphi),
\end{equation}
using the fact that $\varphi$ satisfies Eq.(\ref{schro_time_electric}), the previous equality can be written as 
\begin{equation}\label{}
i\hbar\frac{\partial({\hat E}\varphi)}{\partial t}={\hat H}({\hat E}\varphi),
\end{equation}
it proves that 
\begin{equation}\label{}
{\hat E}\varphi(x,t)= \left(\frac{q^{2}{\cal E}^{2}}{2m}t^{2}-q{\cal E}x\right)\varphi
\end{equation}
is another solution of Eq.(\ref{schro_time_electric}). This characteristic can be generalized to any $j\in\mathbb{N}$ applications of the operator ${\hat E}$, since $[{\hat E}^{j},{\hat H}]=0$ one can prove that 
\begin{equation}\label{}
i\hbar\frac{\partial({\hat E}^{j}\varphi)}{\partial t}={\hat H}({\hat E}^{j}\varphi).
\end{equation}
Hence, we have a countable set of solutions. Defining each one of these functions as $\varphi^{j}(x,t)={\hat E}^{j}\varphi$ such that $\varphi^{0}=\varphi$, the following commutator can be calculated 
\begin{equation}\label{}
[{\hat f},{\hat E}^{j+1}]=i\hbar q{\cal E} (j+1){\hat E}^{j},
\end{equation}
and applying it to $\varphi$ we have that each one of these functions satisfies the following eigenvalue equation
\begin{equation}\label{}
{\hat f}{\hat E}\varphi^{j}=i\hbar q{\cal E} (j+1)\varphi^{j}.
\end{equation}
It is important to note that the eigenvalues for the above equation are imaginary. Despite the fact that the operators in Eq.(\ref{f}) and Eq.(\ref{E}) are Hermitian on their own, the product of two Hermitian operators is not necessarily Hermitian. Hence, ${\hat f}{\hat E}$ can have imaginary eigenvalues.

The general solution of this system can be written, as a superposition of the wave function Eq.(\ref{fundamentalsol1}) and its degeneration, that is
\begin{equation}\label{}
\Psi(x,t)=\sum_{j=0}^{\infty}c_{j}{\hat E}^{j}\varphi(x,t).
\end{equation}
where $c_{j}$ are constants. Due to the degeneracy, it can be difficult to work with the general solution, but there is a specific selection of constants in the linear combination that can simplify the wave function. If the constants are written as
\begin{equation}\label{}
c_{j}=\frac{1}{j!}\frac{\delta t^{j}}{(i\hbar)^{j}},
\end{equation}
the general solution can be written as 
\begin{equation}\label{genaral_sol_electric}
\Psi(x,t)=\varphi(x,t-\delta t),
\end{equation}
where $\varphi(x,t-\delta t)=\psi(x,t)$ defined by Eq.(\ref{psi_xt_electric}). Even though one might assume that the general solution can be obtained straightforwardly by solving Eq.(\ref{eigen_value_electric}) and Eq.(\ref{schro_C}), it is important to note that this is just a particular case of the general solution. The possibility that different selections of constants $c_{j}$ will lead to different effects of the degeneracy must not be discarded. 

Finally, an important property of this system can be deduced when the wave function in Eq.(\ref{genaral_sol_electric}) is invariant under a unitary transformation of the operator Eq.(\ref{Uf}), that is  
\begin{equation}\label{}
{\hat U}_{x}\Psi(x,t)=\exp\left(i\frac{q{\cal E}}{\hbar}\delta x	\delta t\right)\Psi(x,t),
\end{equation}
which leads to the condition that 
\begin{equation}\label{quan_elect_cond}
\frac{q{\cal E}}{\hbar}\delta x\delta t=2\pi n,\quad n\in\mathbb{N}.
\end{equation}
To interpret the above quantity, we rewrite it as follows 
\begin{equation}\label{}
\frac{q^{2}}{\hbar}{\cal E}\delta x\frac{\delta t}{q}=2\pi n,
\end{equation}
then we note that $q^{2}/\hbar$ is the inverse of Klitzing's constant \cite{klitzing1980new}, ${\cal E}\delta x$ is the voltage, and $q/\delta t$ is the current of a single particle. Therefore, redefining Plank's constant as $h=2\pi\hbar$ and rearranging 
\begin{equation}\label{resistance_quant}
R=\frac{h}{q^{2}} n,
\end{equation}
where $R$ is the resistance of the system. This means that the resistance produced by the particle is quantized as integer multiples of the von Klitzing constant.

The quantized resistance derived above exhibits noteworthy characteristics. It is directly proportional to an integer, reminiscent of the quantization observed in the FQHE, first measured by Tsui, Störmer, and Gossard \cite{tsui1982two}. It is well established that the FQHE arises from strong electron-electron interactions in a two-dimensional electron gas subjected to a strong magnetic field \cite{ju2024fractional}. Nevertheless, exploring alternative mechanisms for realizing FQHE-like quantized resistance remains an active area of research \cite{wu2024time,wang2024realization,pu2024microscopic,agarwal2025fractional,wang2025fractional,wang2025developing}, particularly in graphene and multilayer graphene lattices \cite{an2025fractional,lu2024fractional,balram2024fractional,kang2024evidence,xie2024integer,han2024large,xie2024even,zhou2024fractional,das2024zero,kousa2025orbital,zhang2025excitons,huang2025fractional,kim2025observation}.
In contrast with the electron-electron interaction FQHE hypothesis, the quantization eq.(\ref{resistance_quant}) emerges from a single-particle model confined to one dimension, without invoking many-body effects or the presence of a magnetic field. This highlights the fundamentally different physical origin of the quantization observed here.\\

To finalize, an analogy with classical mechanics can be made. Using the wave function in eq. (\ref{fundamentalsol1}) and substituting it into the definition of the quantum current,  
\begin{equation}
    J = \frac{i\hbar}{2m} \left( \varphi\frac{\partial\varphi^{*}}{\partial x} - \varphi^{*}\frac{\partial\varphi}{\partial x}\right),
\end{equation}
one can obtain 
\begin{equation}\label{current_newton}
    J = \frac{q{\cal E}t}{m} |\varphi|^{2}.
\end{equation}
Since the current is, by definition, the velocity $v$ times the particle density $\rho = |\varphi|^{2}$, i.e., $J = v \rho$, from the above equation, the velocity is given by  
\begin{equation}\label{vel_newton}
    v = \frac{q{\cal E}t}{m},
\end{equation}
and, differentiating with respect to time and multiplying by the mass, one can recover the classical Newton force for a constant electric field:
\begin{equation}\label{vel_newton}
    m \frac{dv}{dt} = q{\cal E}.
\end{equation}


\section{Constant electromagnetic field}
As mentioned in the introduction, another autonomous system with a time-dependent conserved operator is the case of a constant electromagnetic field. The analysis for the situation where the electric field is perpendicular to the magnetic field can be found in reference \cite{lizarraga2024quantization}. Here, it will be consider the scenario where the electric field is parallel to the magnetic field.
In this case, the same electric potential than in the previous section is used. On the other hand, to consider the magnetic field in Hamiltonian Eq.(\ref{H}), it is necessary to define the vector potential. In this case ${\bf A}=B(0,0,y)$ satisfies the requirement that the magnetic field is parallel to the electric field. This selection of potentials leads to the following Hamiltonian
\begin{equation}\label{HBPE}
{\hat H}=\frac{1}{2m}\left\{{\hat p}_{x}^{2}+{\hat p}_{y}^{2}+({\hat p}_{z}-m\omega_{c}y)^{2}\right\}-q{\cal E}x
\end{equation}
where cyclotron frequency $\omega_c=qB/mc$ was defined. This hamiltonian has the following conserved operators 
\begin{equation}\label{f2}
{\hat \pi_{x}}={\hat p}_{x}-q{\cal E}t,
\end{equation}
\begin{equation}\label{py2}
{\hat \pi_{y}}={\hat p}_{y}-m\omega_{c}z,
\end{equation}
\begin{equation}\label{pz}
{\hat \pi_{z}}={\hat p}_{z},
\end{equation}
and
\begin{equation}\label{E2}
{\hat E}=i\hbar\frac{\partial }{\partial t}.
\end{equation}
The Schr\"odinger equation can be written as 
\begin{equation}\label{SE2}
i\hbar\frac{\partial\psi}{\partial t}={\hat H}\psi,
\end{equation}
where $\psi=\psi(x,y,z,t)$. For this specific situation, it is helpful to write down the wave function as the product of two time-dependent functions as follows
\begin{equation}\label{basis2}
\psi(x,y,z,t)=\phi_{1}(x,t)\phi_{2}(y,z,t),
\end{equation}
substituting in the Schr\"odinger Eq.(\ref{SE2}), it leads to next couple of equations 
\begin{equation}\label{P2E1}
i\hbar\frac{\partial\phi_{1}}{\partial t}=\frac{1}{2m}{\hat p_{x}}^{2}\phi_{1}-q{\cal E}x\phi_{1},
\end{equation}
and 
\begin{equation}\label{P2E2}
i\hbar\frac{\partial\phi_{2}}{\partial t}=\frac{1}{2m}\left\{{\hat p}_{y}^{2}\phi_{2}+({\hat p}_{z}-m\omega_{c}y)^{2}\phi_{2}\right\}.
\end{equation}
The equation (\ref{P2E1}) is exactly the same than the one solved in the previous section, Eq.(\ref{schro_time_electric}). Therefore, the previous analysis is valid and the wave function $\phi_{1}$ can be written as Eq.(\ref{fundamentalsol1})
\begin{equation}\label{}
\phi_{1}(x,t)=\frac{1}{L}\exp\left(-i\frac{q^{2}{\cal E}^{2}}{6m\hbar}t^{3}+i\frac{q{\cal E}}{\hbar}tx\right).
\end{equation}
On the other hand, the solutions for the Eq.(\ref{P2E2}) can be found with the help of the conserved operators Eq.(\ref{py2}) and Eq.(\ref{pz}). First, we proceed by separating the time coordinate as usual, by 
\begin{equation}\label{}
\phi_{2}(y,z,t)=\exp\left(-i\frac{E}{\hbar}t\right)\overline\phi_{2}(y,z),
\end{equation}
leading to the eigenvalue equation 
\begin{equation}\label{eigenvalueseq2}
E\overline\phi_{2}=\frac{1}{2m}\left\{{\hat p}_{y}^{2}\overline\phi_{2}+({\hat p}_{z}-m\omega_{c}y)^{2}\overline\phi_{2}\right\},
\end{equation}
where $E$ is the energy of the system.
Then, one can use these operators to write down a couple of eigenvalue equations and similar to the previous case the eigenvalues have the effect to displace the solutions by a quantity denoted as $\delta y$ and $\delta z$, such that allow us to write down the equations as 
\begin{equation}\label{eigen21}
{\hat p}_{y}\overline\phi_{2}-m\omega_{c}z\overline\phi_{2}=m\omega_{c}\delta z\overline\phi_{2},
\end{equation}
\begin{equation}\label{eigen22}
{\hat p}_{z}\overline\Phi_{2}=m\omega_{c}\delta y\overline\Phi_{2},
\end{equation}
where $\delta z$ and $\delta y$ are constants having length as units. Both expressions will lead to valid solutions of the system. The simplest one is Eq.(\ref{eigen22}), it leads to a solution of the form
\begin{equation}\label{separable}
\overline\Phi_{2}=\exp\left(i\frac{m\omega_{c}}{\hbar}z\delta y \right){\cal D}(y),
\end{equation}
where ${\cal D}(y)$ is a function to be determined by substituting in Eq.(\ref{eigenvalueseq2}). Hence, this function satisfies the equation 
\begin{equation}\label{}
E{\cal D}=\frac{1}{2m}\left\{{\hat p}_{y}^{2}{\cal D}+m^{2}\omega_{c}^{2}(y-\delta y)^{2}{\cal D}\right\}.
\end{equation}
This equation can be identify as the displaced harmonic oscillator. Therefore, 
\begin{equation}\label{}
{\cal D}(y)=\varphi_{n}\left(\sqrt{\frac{m\omega_{c}}{\hbar}}(y-\delta y)\right),
\end{equation}
where
\begin{equation}\label{harmOsci}
\varphi_{n}(\xi)=\frac{1}{\sqrt{2^{n}n!}}\left(\frac{m\omega_{c}}{\pi\hbar}\right)^{1/4}\exp(-\xi^{2})H_{n}(\xi),
\end{equation}
such that $H_{n}$ are the Hermite polynomials and the spectrum is given by the Landau's levels
\begin{equation}\label{LandauL}
E_{n}=\hbar\omega_{c}\left(n+\frac{1}{2}\right).
\end{equation}
This procedure is basically the application of Landau's {\it ansatz}. However, there is an important mathematical detail to be considered. The expression Eq.(\ref{separable}), is a separable variable solution of the form $f(z)g(y)$, but if one attempts to formally separate the Eq.(\ref{eigenvalueseq2}) it can be notice that in fact the equation can not be separated. This detail gives and insight that the solution of the Eq.(\ref{eigen21}) is still needed. From Eq.(\ref{eigen21}), one obtains that 
\begin{equation}\label{segundaSolucion}
\overline\phi_{2}=\exp\left(i\frac{m\omega_{c}}{\hbar}y(z-\delta z)\right){\cal C}(z),
\end{equation}
where ${\cal C}(z)$ is a function to be determined . By substituting this last expression in Eq.(\ref{eigenvalueseq2}) one obtains the following equation 
\begin{equation}\label{}
E{\cal C}=\frac{1}{2m}\left\{\hat{p}_{z}^{2}{\cal C}+m^{2}\omega_{c}^{2}(z-\delta z)^{2}{\cal C}\right\},
\end{equation}
once again, the displaced harmonic oscillator equation is obtained. Hence, the solutions are 
\begin{equation}\label{}
{\cal C}(z)=\varphi_{n}\left(\sqrt{\frac{m\omega_{c}}{\hbar}}(z-\delta z)\right),
\end{equation}
where $\varphi_{n}$ is given by the expression Eq.(\ref{harmOsci}), having the Landau levels as eigenvalues Eq.(\ref{LandauL}) .\\

At this point, is helpful to use the conserved operators to define a set of unitary operators 
\begin{equation}\label{Ux}
{\hat U}_{x}=\exp\left(-i\frac{\delta x}{\hbar}({\hat p}_{x}-q{\cal E}t)\right),
\end{equation}
\begin{equation}\label{Uy}
{\hat U}_{y}=\exp\left(-i\frac{\delta y}{\hbar}({\hat p}_{y}-m\omega_{c}z)\right),
\end{equation}
\begin{equation}\label{Uz}
{\hat U}_{z}=\exp\left(-i\frac{\delta z}{\hbar}{\hat p}_{z}\right),
\end{equation}
\begin{equation}\label{Ut}
{\hat U}_{t}=\exp\left(i\frac{\delta t}{\hbar}{\hat E}\right).
\end{equation}
Such that they define the symmetries of the system
\begin{equation}\label{}
{\hat H}-{\hat E}={\hat U}_{i}^{\dagger}({\hat H}-{\hat E}){\hat U}_{i},\quad i=x,y,z,t.
\end{equation}
They can be used to rewrite the wave functions Eq.(\ref{separable}) and Eq.(\ref{segundaSolucion}) as a unitary transformation of their respective harmonic oscillator, that is 
\begin{equation}\label{}
\overline\Phi_{2}={\hat U}_{y}\varphi_{n}\left(\sqrt{\frac{m\omega_{c}}{\hbar}}y\right),
\end{equation}
and
\begin{equation}\label{}
\overline\phi_{2}={\hat U}_{z}\exp\left(i\frac{m\omega_{c}}{\hbar}yz\right)\varphi_{n}\left(\sqrt{\frac{m\omega_{c}}{\hbar}}z\right).
\end{equation}
This observation is useful to define a couple of time dependent wave functions using the respective inverse unitary transform
\begin{equation}\label{}
\zeta_{n}(x,y,t)=\exp\left(-i\frac{E_{n}}{\hbar}t\right)\phi_{1}{\hat U}_{y}^{\dagger}\overline\Phi_{2},
\end{equation}
and 
\begin{equation}\label{}
\overline\zeta_{n}(x,y,z,t)=\exp\left(-i\frac{E_{n}}{\hbar}t\right)\phi_{1}{\hat U}_{z}^{\dagger}\overline\phi_{2},
\end{equation}
both of the above functions still have the Landau's levels as eigenvalues of the Hamiltonian Eq.(\ref{HBPE}). 

Before writing down the general solution, it is important to consider the degeneration of the system. At the beginning of this work the expression Eq.(\ref{expr1}) was obtained for a time dependent operator. However, for the particular case when the operator is time-independent, for instance $\hat{\pi}_{y}$, the following can be written when the equality is applied to $\zeta_{n}$ 
\begin{equation}\label{}
{\hat H}(\hat{\pi}_{y}\zeta_{n})=\hat{\pi}_{y}({\hat H}\zeta_{n})=E_{n}(\hat{\pi}_{y}\zeta_{n}).
\end{equation}
Similarly, for the operator $\hat{\pi}_{z}$ and the function $\overline\zeta_{n}$ the following equality is obtained 
\begin{equation}\label{}
{\hat H}(\hat{\pi}_{z}\overline\zeta_{n})=\hat{\pi}_{z}({\hat H}\overline\zeta_{n})=E_{n}(\hat{\pi}_{z}\overline\zeta_{n}).
\end{equation}
This last two equalities gives the insight that $\hat{\pi}_{y}\zeta_{n}$ and $\hat{\pi}_{z}\overline\zeta_{n}$ are also solutions of Hamiltonian Eq.(\ref{HBPE}) having the Landau's levels as eigenvalues, even though if after the application of the respective operator the wave function obtained is not proportional to $\zeta_{n}$ or $\overline\zeta_{n}$. 
This observation can be easily generalized to any $j\in\mathbb{N}$ applications of the operators, which means that $\hat{\pi}_{y}^{j}\zeta_{n}$ and $\hat{\pi}_{z}^{j}\overline\zeta_{n}$ are also eigenfunctions of the system. On the other hand, the situation with the time-dependent operator $\hat{\pi}_{x}$ is not so different. Once again, from Eq.(\ref{expr1}) applied to the function $\zeta_{n}$, which satisfies the time-dependent Schr\"odinger equation 
\begin{equation}\label{}
\hat{H}(\hat{\pi}_{x}\zeta_{n})=i\hbar\frac{\partial (\hat{\pi}_{x}\zeta_{n})}{\partial t},
\end{equation}
similarly for $\overline\zeta_{n}$
\begin{equation}\label{}
\hat{H}(\hat{\pi}_{x}\overline\zeta_{n})=i\hbar\frac{\partial (\hat{\pi}_{x}\overline\zeta_{n})}{\partial t}.
\end{equation}
It can be noted that $\hat{\pi}_{x}\zeta_{n}$  and $\hat{\pi}_{x}\overline\zeta_{n}$ are also solutions of the Schr\"odinger equation. This observation can be generalized to $j$ applications of the same operators and one can realize that $\hat{\pi}_{x}^{j}\zeta_{n}$ and $\hat{\pi}_{x}^{j}\overline\zeta_{n}$ are also solutions of the Schr\"odinger equation. A final important observation is that the wave functions obtained due to this part of the degeneration are time dependent and since they are not proportional to the original functions $\zeta_{n}$ or $\overline\zeta_{n}$ the aplication of the operator $\hat{E}$ will also give a new solution of the Scr\"odinger equation and also will be the case for $j$ applications of the operator. 
Thus, the general solution, $\Psi$, is the linear combination of the solutions considering the degeneration of the system
\begin{equation}\label{}
\Psi=\sum_{n,j,j',k}a_{njj'k}\hat{E}^{k}\hat{\pi}_{x}^{j}\hat{\pi}_{y}^{j'}\zeta_{n}+\sum_{n,j,j',k}\overline{a}_{njj'k}\hat{E}^{k}\hat{\pi}_{x}^{j}\hat{\pi}_{z}^{j'}\overline\zeta_{n},
\end{equation}
where $a_{njj'k}$ and $\overline{a}_{njj'k}$ are constants. 

This last expression could be difficult to address analytically. However, there is a particular case for the constant which is worth to mention here. Being the constants 
\begin{equation}\label{LG_Constants_1}
a_{n,j,j',k}=a_{n}\frac{1}{k!}\frac{\delta t^{k}}{(i\hbar)^{k}}\frac{(-1)}{j!}\frac{\delta x^{j}}{(i\hbar)^{j}}\frac{(-1)}{j'!}\frac{\delta y^{j'}}{(i\hbar)^{j'}},
\end{equation}
and
\begin{equation}\label{LG_Constants_2}
\overline{a}_{n,j,j',k}=\overline{a}_{n}\frac{1}{k!}\frac{\delta t^{k}}{(i\hbar)^{k}}\frac{(-1)}{j!}\frac{\delta x^{j}}{(i\hbar)^{j}}\frac{(-1)}{j'!}\frac{\delta y^{j'}}{(i\hbar)^{j'}},
\end{equation}
they lead to the following for of the general solution 
\begin{equation}\label{}
\Psi=\sum_{n}a_{n}{\hat U}_{t}{\hat U}_{x}{\hat U}_{y}\zeta_{n}+\sum_{n}\overline{a}_{n}{\hat U}_{t}{\hat U}_{x}{\hat U}_{z}\overline\zeta_{n}.
\end{equation}
If one propose this wave function to be invariant under the action of the unitary operator ${\hat U}_{x}$, meaning that ${\hat U}_{x}\Psi=\Psi$, one can realize that the following equality is obtained  
\begin{equation}\label{}
{\hat U}_{x}\Psi=\exp\left(i\frac{q{\cal E}}{\hbar}\delta x\delta t\right)\Psi,
\end{equation}
implying, once again, that the invariance of the wave function under this unitary transformation implies the quantization of resistance Eq.(\ref{resistance_quant}).




\section*{Conclusions}

In this work, it was demonstrated that time-dependent conserved operators can be used to solve the Schrödinger equation for two autonomous systems, applying the Lewis–Riesenfeld invariant method. The degeneracy of the systems was explored, revealing how these conserved operators act as generators of solutions to the corresponding partial differential equation and how they define the symmetries of the systems. The invariance of the wave function under the unitary transformation  $\hat{U}_x$ leads to resistance quantization, expressed as integer multiples of the von Klitzing constant.

The condition \( \hat{U}_x \Psi = \Psi \) suggests a scenario in which all electrons in a closed system initially occupy the same quantum state. After a displacement by \( \delta x \), the wave function returns to its initial form, demonstrating that this symmetry is necessary for quantization.

Finally, the quantized resistance given by Eq.~(\ref{resistance_quant}) closely resembles the form observed in the Störmer experiment. While the FQHE is commonly attributed to strong electron-electron interactions in two-dimensional systems under high magnetic fields, the results presented here show that similar resistance quantization can arise in much simpler, non-interacting single-particle models. This emphasizes that symmetry and conserved quantities alone can lead to quantized behavior reminiscent of the FQHE—even in the absence of interactions and, in the first system, under the influence of an electric field only.

\section*{Acknowledgment}
This work was supported by DGAPA-UNAM Posdoctoral Program (POSDOC), the project UNAM-PAPIIT IG100725 and project SECIHIT Fronteras CF-2023-G. I am grateful for these supports.

\newpage
\bibliography{bibliografia}

\end{document}